\documentclass[
nopreprint%choose either nopreprint or preprint
]{jasatex}

\usepackage{color}   % added by ANN
\usepackage{graphicx}% Include figure files
\usepackage{dcolumn}% Align table columns on decimal point
\usepackage{amsmath,amsfonts,amssymb}% popular packages from the American Mathematical Society
\usepackage{bm}% Bold Math package

\newcommand{\beq}[1]{  \begin{equation} \label{#1} }  
\newcommand{\eeq}{     \end{equation}} 	
\newcommand{\bal}[1]{  \begin{align} \label{#1} }  
\def\dd{\operatorname{d}}

\begin{document} %%%%%%%%%%%%%%%%%%%%%%%%%%%%%%%%%%%%%%%%%%%%%%%%%%%%%%%%%%%%%%%%%%%%%%%%

\title{
\textcolor{blue}{Acoustic metafluids made from  three acoustic fluids  }  
}

\author{Andrew N. Norris \, \&  Adam J. Nagy}
  \email{norris@rutgers.edu}
\affiliation{Mechanical and Aerospace Engineering, Rutgers University, Piscataway NJ 08854}

\date{\today}% It is always \today, today,  but any date may be explicitly specified

\begin{abstract}
Significant reduction in target strength and radiation signature can be achieved by surrounding an object with multiple concentric layers comprised of three acoustic fluids.  The idea is to make a  finely layered shell with the thickness of each layer defined by a unique transformation rule.  The shell has the effect of steering incident acoustic energy around the structure, and conversely, reducing the radiation strength.  The overall effectiveness and the precise form of the layering depends upon the  densities and compressibilities of the three fluids.  Nearly optimal results are obtained if one fluid has density equal to the background fluid, while the other two densities are much greater and much less than the background values.  Optimal choices for the compressibilities are also found.   Simulations in 2D and 3D illustrate  effectiveness of the three fluid shell. 
The limited range of acoustic metafluids that are possible using only two fluid constituents is also discussed. 

\end{abstract}

\maketitle

\section{Introduction}\label{sec1}

The idea behind transformation acoustics is that  a coordinate transformation makes it possible to have one region of an acoustic fluid mimic another region.   Fluids that have this property have been called acoustic metafluids.     In transformation optics the transformation uniquely defines the material properties, but this is not the case in acoustics, and  there is an added degree of freedom in the  makeup of the acoustic metafluid.  The range of possible acoustic metafluids has been derived \cite{Norris09}, and includes fluids with anisotropic inertia  and pentamode materials.

Interest in transformation acoustics has been motivated by the 
possibility of acoustic cloaking.  The first electromagnetic wave cloaking device \cite{Pendry06} uses transformation of coordinates in the governing wave equation to steer energy around the cloaked object.  It was subsequently  demonstrated  that the same methods should work for the acoustic wave equation \cite{Cummer07,Chen07}.  The acoustic cloak corresponds to the  limiting  case of a point transformed into a finite region, and it has  unavoidable physical singularities associated with the extreme nature of the transformation.  Different types of singularities are obtained depending on whether the transformed metafluid is purely inertial with anisotropic density and a single bulk modulus, or in the other limit,  purely pentamodal with isotropic inertia. 
The distinction is important for cloaking, for which  it is known that use of only   fluids with anisotropic inertia (inertial cloaks) requires infinite mass, and is therefore not a realistic path towards acoustic cloaking \cite{Norris08b}.  Despite this limitation, it is possible to achieve almost perfect, or near-cloaking, using layers of anisotropic fluids that approximate the transformed medium, without the singularity.  
%Examples of this type of layering have been proposed  \cite{Torrent08b,Scandrett10}.   
For instance,  Torrent and S\'anchez-Dehesa \cite{Torrent08b} partition the shell into many small but equally thin  layers where the local properties are defined by two normal fluids, with density and bulk moduli  $\{ \rho_j, K_j\}$, $j=1,2$, such that the  averaged quantities  
$\rho_r = \frac12 (\rho_1 + \rho_2)$, 
$\rho_{\perp} =[\frac12 (\rho_1^{-1} + \rho_2^{-1})]^{-1}$ and  $K = [\frac12 (K_1^{-1} + K_2^{-1})]^{-1}$
yield the anisotropic metafluid properties 
$\{ \rho_r (r), 
 \rho_{\perp} (r), K(r)\}$ 
proposed by Cummer and Schurig \cite{Cummer07}.   In order to achieve this equivalence it is necessary to make  $\{ \rho_j, K_j\}$, $j=1,2$, functions of $r$, with the result   a large number of  distinct fluids is necessary:   100 and 400 for the two numerical examples reported by Torrent and S\'anchez-Dehesa  \cite{Torrent08b}.   

The purpose of this paper is to demonstrate that significant reduction in target strength can be achieved using layers comprised of only three acoustic fluids.  The idea is to make a  finely layered shell that surrounds the structure, with each layer being one of the three fluids, but instead of prescribing the relative thickness of each layer we allow it to be a function of $r$.  
The transformation formulas then imply  unique values  for the relative concentrations as functions of $r$, in both two (cylinder) and three (sphere) dimensions.  

The outline of the paper is as follows.  The homogenized layered shell  and the transformation metamaterial are introduced separately in  Section \ref{sec2} in the context of an $N-$fluid material.   The remainder of the paper concentrates on the 3-fluid $(N=3)$  configuration.  General results for both  cylindrical and spherical shells are derived in Section \ref{sec4}, including the unique transformation formulae.  Dependence of the cylindrical transformation metamaterial on the constituent properties of the 3-fluids is explored in Section \ref{sec5}.  The explicit nature of the transformation formulae for 2D  suggest optimal choices for the fluid densities and compressibilities.  These findings are confirmed in 
Section \ref{sec6} where examples of cylindrical and spherical 3-fluid metamaterials are presented. 
Numerical simulations showing their effectiveness in reducing  scattering strength
  in 2 and 3 dimensions are also presented in Section \ref{sec6}.
%The 2-fluid $(N=2)$ material is shown in Section \ref{sec3} to be too restrictive to generate a useful transformation metamaterial.

\section{Preliminaries}\label{sec2}
 We consider radially symmetric  configurations, cylindrical in 2D and spherical in 3D. 
A fluid annulus or shell occupies $0< r_0\le r \le r_{out}$, and is surrounded by a uniform acoustic medium with density and sound speed $\rho_{out}$, $c_{out}$, in $r>r_{out}$.   The shell is assumed to be made of a finite number, $N$, 
of distinct fluids arranged in a well defined stratification that results in an effective material with smoothly varying properties in the radial direction. We are particularly interested in finding the smallest number $N$  for which it is possible that the  stratification has the properties of  an  acoustic metafluid.   An acoustic metafluid is defined here as a material with 
desirable effective properties that cannot easily be obtained with a single, physical fluid.  This definition obviously includes materials  
obtained by a coordinate transformation of a larger region of  uniform acoustic fluid with properties equal to those of the exterior fluid in   $r>r_{out}$.  
%Clearly, $N= 1$ fluid is insufficient, and we therefore concentrate on examining the cases $N=2$ and $N=3$.  

For simplicity, but with no lack in generality, we set
 $r_{out} =1$, $c_{out} =1$ and $\rho_{out} =1$,  which is equivalent to choosing units for length, time and mass, respectively.  For the remainder of the paper all quantities are non-dimensional. 
 
  We first consider the 
 homogenized shell   composed of a   layering of $N$  distinct fluids defined by their mass densities, $\rho_1, \ldots, \rho_N$, and the compressibilities $C_1, \ldots, C_N$. The compressibility is $C_i=K_i^{-1}$ where $K_i$ is the bulk modulus, and the  
  wave speeds are  $c_i = \sqrt{K_i/\rho_i}$, and the impedances are 
  $z_i = \sqrt{K_i \rho_i }$, $i=1,\ldots, N$. We define, for later use, 
$ S_i = \rho_i C_i$, or alternatively, 
  $S_i =  c_i^{-2}$, so that we may  identify $\sqrt{S_i}$ as  acoustic slowness in fluid $i$.

  The layering yields an effective fluid with compressibility $C_*$ and  anisotropic inertia  defined by radial density $\rho_r$, and circumferential density $\rho_\perp$.  The parameters of the effective fluid are defined by  homogenization of the stratified medium as  \cite{Schoenberg83} 
\beq{4-3}
\begin{pmatrix}
\rho_r 
\\ 
\rho_\perp^{-1}
\\ 
C_* 
\end{pmatrix}
=
\begin{pmatrix}
\langle \rho \rangle
\\ 
\langle \rho^{-1} \rangle
\\ 
\langle C \rangle
\end{pmatrix} , 
\eeq
where $\langle \cdot \rangle$ is the local average over the volume fractions of the $N$-fluids,
\beq{-54}
\langle x\rangle = \sum_{i=1}^N \phi_i x_i , 
\quad\text{with  }\langle 1 \rangle = 1 .
\eeq
It is assumed that $\phi_i = \phi_i (r)$, so  that the averages \eqref{4-3} define parameters $\rho_r(r)$,  $\rho_\perp(r)$, and $C_*(r)$.  This type of inhomogeneous  or localized homogenization  may be achieved by allowing the layering to be sufficiently fine, and  will be illustrated by numerical examples later.

The  transformation from the current (physical) domain to the mimicked one makes the shell appear acoustically as if it is a larger shell of fluid with uniform properties equal to the exterior fluid.  The key is a transformation function, $r\rightarrow R = R(r)$, such  that the range of $R$ exceeds its domain, i.e.,  the inverse mapping $R\rightarrow r$ physically contracts space. To be specific, the outer boundary is mapped to itself, $r=R=1$, and the inner boundary  $r=r_0$ is mapped to  $R=R_0$, with $0< R_0 = R(r_0) <r_0$. The perfect acoustic cloak is defined by $R_0 =0$.  
The transformed  material has  properties    $\rho_{rT}$,  $\rho_{\perp T}$, and $C_{*T}$, with  values  uniquely  defined by  the transformation in $d-$dimensions 
 as \cite{Norris08b}
\beq{4-4}
\begin{pmatrix}
\rho_{rT}
\\ 
\rho_{\perp T}^{-1}
\\ 
C_{*T} 
\end{pmatrix}
= R' 
\begin{pmatrix}
(r/R)^{d-1}
\\ 
(r/R)^{3-d}
\\ 
(r/R)^{1-d}
\end{pmatrix} , \quad d=2\text{ or }3,
\eeq
and where $R' = \dd R /\dd r$. 

The connection between the homogenized material \eqref{4-3} and the acoustically transformed material  \eqref{4-4} is now made explicit by requiring 
$\rho_{rT} = \rho_{r}$, $\rho_{\perp T}= \rho_\perp $ and $C_{*T}=C_*$ (and we drop the subscript $T$).  Our objective is to find families of transformation functions $R= R(r)$,  $\phi_i = \phi_i (r)$ for which this equivalence can be achieved.  It depends, of course, on the choices of material properties $\{\rho_i, C_i\}$, $i=1,\ldots ,N$, and not all combinations will work.   Among the requirements are  that the transformation function   is one-to-one, and that the volume fractions are all  between zero and unity.  
We therefore require that 
${\pmb \phi} \in \Phi_N$ where  ${\pmb \phi}$ is the $N-$dimensional vector of volume fractions, and $\Phi_N$ the   $N-1$ dimensional  surface   on which it must lie,
\beq{03-}
\Phi_N = \{\phi_i \ge 0, \,\,  \sum_i \phi_i =1,\,\, i=1,\ldots , N\}.  
\eeq
%In order to gain  some understanding of the problem we start with the simpler case $N=2$ and then move on to consider $N=3$. 

 \section{The three fluid material}\label{sec4}

% The extra fluid adds a significant amount of freedom in that we do not expect $\rho_r = \rho_\perp^{-1}$, $C_*$ and the volume fractions  to be constrained to single values.  

\subsection{Algebraic formulation}

The first two relations in  \eqref{4-3} and the identity $\eqref{-54}_2$ may be written in matrix form for   $N=3$,   
\beq{1}
\begin{pmatrix}
1 & 1 & 1
\\
 \rho_1 & \rho_2  & \rho_3
\\
 {\rho_1^{-1}} &    {\rho_2^{-1}}  
&   {\rho_3^{-1}}
\end{pmatrix}
\begin{pmatrix}
\phi_1 \\ \phi_2 \\ \phi_3
\end{pmatrix}
= \begin{pmatrix}1
\\
 \rho_r
\\
 \rho_\perp^{-1}
\end{pmatrix} . 
\eeq
This can be solved to give  the 3-vector of volume fractions in terms of 
$\rho_r$ and $\rho_\perp^{-1}$.  Substitution  into the third relation in  \eqref{4-3} yields an expression for $C_*$ in terms of 
$\rho_r$ and $\rho_\perp^{-1}$.  Thus, 
 \begin{subequations}
\bal{40}
{\pmb \phi} &= {\pmb f}_0 + 
\rho_r {\pmb f}_1 + 
\rho_\perp^{-1} {\pmb f}_2 
,
\\
C_{*} &= \alpha + \beta_1 \rho_r + \beta_2 \rho_\perp^{-1}, \label{69}
\end{align}
 \end{subequations}
where the 3-vectors in \eqref{40} are 
\bal{41}
{\pmb \phi} &=
\begin{pmatrix}
\phi_1 \\ \phi_2 \\ \phi_3
\end{pmatrix}
 ,%\quad 
& {\pmb f}_0   =D
\begin{pmatrix}
\frac{\rho_2}{\rho_3} - \frac{\rho_3}{\rho_2} \\  
\frac{\rho_3}{\rho_1} - \frac{\rho_1}{\rho_3} \\  
\frac{\rho_1}{\rho_2} - \frac{\rho_2}{\rho_1}
\end{pmatrix}
 ,\nonumber \\  
 {\pmb f}_1 &=D
 \begin{pmatrix}
\frac1{\rho_2}-\frac1{\rho_3} \\  
\frac1{\rho_3}-\frac1{\rho_1} \\   
\frac1{\rho_1}-\frac1{\rho_2}
\end{pmatrix}
  , 
  %\quad 
  &{\pmb f}_2 =D
  \begin{pmatrix}
\rho_3-\rho_2 \\  
\rho_1-\rho_3 \\  
\rho_2-\rho_1
\end{pmatrix}
  , 
  \end{align} %%%%%%%%%%%%%%%%%%%%%%%%%%%%%%%%%%%%%%%%%%%%%%%%%%%%%%%%%%%%%%%%%%%%%
  with  
$
D=  {\rho_1\rho_2\rho_3}/[(\rho_1-\rho_2)(\rho_2-\rho_3)(\rho_3-\rho_1)]$,
and  the scalars $\alpha$, $\beta_1$ and  $\beta_2 $ in \eqref{69} are
\beq{7} 
\quad \alpha  = {\pmb C}^T {\pmb f}_0, 
\quad \beta_1 =  {\pmb C}^T 
{\pmb f}_1 , 
\quad 
\beta_2 =  {\pmb C}^T  {\pmb f}_2 , 
\eeq
with $ {\pmb C}^T  = (C_1 , C_2 , C_3)$.

\subsection{The transformation function}
\subsubsection{Differential equations }

Eliminating $C_*$, $\rho_r$ and $\rho_\perp^{-1}$ from \eqref{69} using the identities \eqref{4-4} yields a differential equation for the 
  transformation function, 
%\beq{10} \big( \frac{R}{r}\big)^{d-1}R'  = \alpha  + \bigg(\beta_1  \big(\frac{r}{R}\big)^{d-1} +\beta_2  \big(\frac{r}{R}\big)^{3-d} \bigg)R', %\qquad d=2\text{ or }3.  
%\eeq
%where $d=2$ or $3$ is the spatial dimension. 
%These   differential equations for $R(r)$   must be supplemented by the condition $R(1)=1$.  They may be rewritten 
\beq{0345}
 \frac{\dd R}{\dd r} =  %\times 
\begin{cases} \alpha
\big(\frac{R}{r} -  \beta  \frac{r}{R}\big)^{-1},  & 2D, 
\\
& 
\\ 
\alpha
\big(\frac{R^2}{r^2} - \beta_1 \frac{r^2}{R^2}- \beta_2\big)^{-1}, & 3D, 
\end{cases} 
\eeq 
subject to the boundary condition $R(1)=1$.  The parameters $\beta $ and, for later use, $\lambda$, $\mu$, are defined 
\beq{0566}
 \beta =\beta_1+\beta_2 ,
 \quad %\text{with  }
\lambda = \alpha +\beta
,\quad
\mu = -\frac\beta\alpha .
 \eeq

\subsubsection{2D solution  }
We first consider the 2D equation $\eqref{0345}_1$.  Let 
$
x=r^2$, $ X  =R^2  $, then 
eq.  $\eqref{0345}_1$ becomes 
\beq{23} X\frac{\dd x}{\dd X} +  \frac\beta\alpha x=\frac{X}\alpha ,
\qquad x(1)=1.
%\quad \Rightarrow \quad \frac{\dd }{\dd g}\big( g^{\frac\beta\alpha} x\big) = \frac{1}\alpha g^{\frac\beta\alpha}.
\eeq
Integrating yields
\beq{251}
%x = \frac{X+ (\lambda - 1)X^{\mu}}{ \lambda} \quad \Leftrightarrow \quad
r = \bigg( \frac{R^2+ (\lambda - 1)R^{2\mu}}{ \lambda}\bigg)^{1/2}.
\eeq
 The 2D transformation function is therefore completely defined by the two  parameters
 $\lambda$ and $\mu$, given in explicit form in \eqref{27}.  
 
 \subsubsection{3D solution }
 
 The 3D equation $\eqref{0345}_2$  becomes, with the change of variable  
$ s = \frac{r}{R}$, 
 \bal{091}
 \frac1{R}\frac{\dd R}{\dd s} &=\frac{
 -\alpha s^2 }{ \beta_1 s^4 +  \alpha s^3+\beta_2 s^2 -1}
 \nonumber \\
 &% = \frac{ \alpha s^2}{ \beta_1 \prod\limits_{i=1}^4 (s -s_i)}
 = 
 \sum_{i=1}^4 \frac{\gamma_i}{ s-s_i} ,
 \quad R(1)=1., 
 \end{align} 
 where the four roots $s_i$ and the coefficients $ \gamma_i $, 
 $i=1,2,3,4$,  are defined by 
 \begin{subequations}
  \bal{092} 
 %(s-s_1) (s-s_2) (s-s_3) (s-s_4)
 \beta_1 \prod\limits_{j=1}^4 (s -s_j)
 &=\beta_1 s^4 +   \alpha   s^3+  \beta_2  s^2 - 1,
 \\
  \gamma_i &= \frac{-\alpha  s_i^2}{\beta_1  \prod\limits_{j\ne i} (s_i-s_j) }.
 \end{align}
  \end{subequations}
 Note that $\sum_i \gamma_i =0$, $\sum_i   s_i = -\alpha/\beta_1$, 
  $\sum_i \gamma_i s_i= -\alpha/\beta_1$, 
 $\sum_i \gamma_i s_i^2=  (\alpha/\beta_1)^2$.  
Integration of \eqref{091} yields 
 \beq{094}
R =  \prod\limits_{i=1}^4 \bigg(\frac{ \frac{r}{R} -s_i}{1 -s_i} 
\bigg)^{ \gamma_i} .
 \eeq
This provides an implicit formula for $R$ and $r$, in terms of the three parameters $\alpha$, $\beta_1$ and $\beta_2$.  Using the fact 
that $1\le s \le s_0$,  where $s_0$ is defined in the next subsection, eq. \eqref{094} gives $R$ as a function of $s$, from which $r = sR$ is obtained. 
 
 \begin{figure}[ht]  %%%%%%%%%%%%%% %%%%%%%%%%%%%% %%%%%%%%%%%%%% %%%%%%%%%%%%%% %%%%%%%%%%%%%%
\begin{center}
 \includegraphics[width=2.5in]{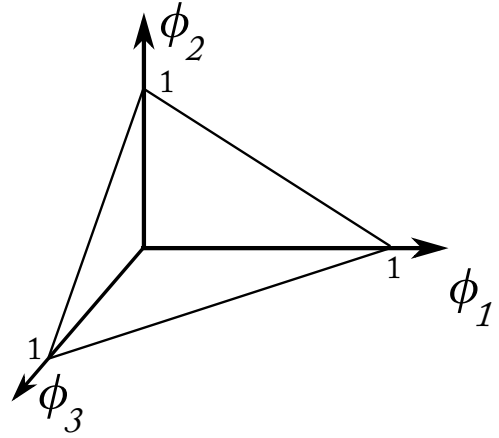} %{phi_map_v_2.eps} 
\end{center} 
\caption{The range of ${\pmb \phi}$ for the 3-fluid.}
 \label{fig1}
\end{figure}

 \subsection{The inner radii $r_0$ and $R_0$}

It  follows from continuity of the solution of the differential equation \eqref{0345}  that the  values of the inner radii $r_0$ and $R_0$ should correspond to a point on the edge of the triangular region $\Phi_3$, see Fig. \ref{fig1}.   The actual radial values can be determined from eq.  \eqref{1},  using $\rho_r$ and $\rho_{\perp}$ as defined in \eqref{4-4} and keeping the parameter $s = \frac{r}{R}$  to express $\phi_i$ of \eqref{40} in the form 
\beq{phijk}
\phi_i = \frac{\rho_i \big[ (s^{d-1} + \rho_j \rho_k s^{3-d} )R' - (\rho_j + \rho_k)\big]
}{(\rho_i-\rho_j)(\rho_i - \rho_k)}, 
\eeq
where $i\ne j\ne k\ne i$.  Replacing $R'$ by \eqref{0345} and setting \eqref{phijk} to zero implies an algebraic (polynomial) equation for $s$.   In principle there are three possible solutions, corresponding to each of $\phi_i =0$, $i=1,2,3$.  However, in practice for a given set of 3-fluids only one is important, and we choose the 3-fluid properties so that it is the root for $\phi_2 =0$. 
We consider first  $d=2$.

In the 2D cylindrical  configuration the equation $\phi_2 =0$  is a quadratic in $s$ with a single positive root greater than unity (corresponding to $r_0>R_0$), which combined with  \eqref{251} implies  
  $R_0$ and $r_0$ in explicit form as     
  \begin{subequations}\label{0632}
\bal{0632a}
R_0 &=   
 \bigg[ ( \lambda -1) 
  \bigg( \frac{  \rho_{r2}^{-1} -\mu}  { 1- \rho_{r2}^{-1}} \bigg)
 \bigg]^\frac1{2(1-\mu )} ,
 \\
r_0 &= 
 \bigg[   \lambda  
  \bigg( \frac{  \rho_{r2}^{-1} -\mu}  { 1- \mu} \bigg)
 \bigg]^{-\frac12 } R_0,
 \end{align}
 \end{subequations}
where 
\beq{-6-6}
\rho_{ri} = \frac{\rho_j+\rho_k }{1+ \rho_j\rho_k}  
\quad (i\ne j\ne k \ne i). 
\eeq

For the 3D spherical case the equation $\phi_2 =0$  becomes a biquadratic in $s$.
% with a single positive root.   
We find 
 \beq{068}
R_0 =    \prod\limits_{i=1}^4 \bigg(\frac{ s_0 -s_i}{1 -s_i} 
\bigg)^{ \gamma_i},
 \quad 
r_0  = s_0R_0 ,
\eeq 
 where $s_0$ is a positive root   of  
\beq{0=3}
s^4[\alpha + \beta_1(\rho_1 + \rho_3)] + s^2[\alpha \rho_1 \rho_3 + \beta_2(\rho_1 + \rho_3)] - (\rho_1 + \rho_3) = 0.
\eeq
The transformation requires $s_0 >1$, and numerical experiments (see Section \ref{sec6}) indicate that a single real root greater than unity exists in all the cases considered. 

 \subsection{Total mass and average density}
The total mass $m$ of the 3-fluid shell is the integral of the local average of the density, $\langle \rho\rangle$.  Therefore, $m$ follows from eq. \eqref{4-3} as the volumetric integral of $\rho_r (r)$.   Substituting from  \eqref{4-4}$_1$ and using \eqref{0345}, the integral can be expressed in closed form for the 2D case, and reduced to an integral in $s=r/R$ for the 3D case.  We find 
\beq{30=}
m  %&\equiv 2\pi \int_{r_0}^1 \rho_r r\dd r 
 =\begin{cases}
  \frac \pi \lambda \big\{ 
1 - R_0^2 + \frac{(\lambda-1)}\mu \big( 1-  R_0^{2\mu}\big)
\big\} , & 2D, 
\\
4\pi \frac{ \alpha}{\beta_1} \int\limits_1^{s_0}  \, s^6
 \prod\limits_{i=1}^4 \frac{ \big(s -s_i\big)^{ 3\gamma_i-1}}{\big(1 -s_i\big)^{ 3\gamma_i}} 
 \dd s, & 3D,
\end{cases}
\eeq
from which the average density in the shell,  $\bar{\rho} = 3m/[\pi (d+1)(1-r_0^d)]$, can be found.  
For 2D we find, after some simplification, 
\beq{31=}
\bar{\rho} = \frac1\mu + \frac1\beta
 \bigg( \frac{1-  R_0^2 }{1-r_0^2 }\bigg) , \quad 2D.
\eeq
 
 \subsection{Summary}
We have shown that the three fluid shell is uniquely related to   possible transformation  functions in both 2- and 3-dimensions.  The connection is still somewhat tentative, since we must confirm that the functions are physically realistic.  This requires among other things that the volume fractions are all positive and between zero and unity, i.e. that ${\pmb \phi} \in \Phi_3$ where  the    equilateral triangle  surface $\Phi_3$ is   defined by  \eqref{03-}.   We must also  confirm that the inner radii are actually given by eqs. \eqref{0632} in 2D and \eqref{068} in 3D.  Optimally, both of the inner radii should be small, since $R_0 \ll 1$
 means that the mapped region $R_0\le R \le 1$ is almost the entire interior of the cylinder/sphere  of radius $1$, while 
 $R_0 \ll r_0$ implies that the shell $r_0 \le r \le 1$ in physical space occupies a relatively small proportion of the mapped region.    
  In the next Section we  consider the  2D shell for which  these questions can be answered in explicit form.  
	
The results for the 3-fluid shell indicate that there are no free parameters for a given set of fluids. This suggests that the transformation property cannot be  
 achieved with only two fluids.  It is shown in Appendix B that the 2-fluid case is too constrained, although it does display some interesting physical properties, even if it cannot provide acoustic cloaking.   

%Before solving for $R$ we note some implications for the properties of the three fluids. 

\section{The three fluid material in $2D$}\label{sec5}

\subsection{Range of material parameters }
The relation  $\rho_r\rho_\perp = 1$, which holds only in 2D (see eq. \eqref{4-4}), considerably simplifies the algebra of the problem as compared with the 3D case, allowing clearer understanding of  parametric dependence.  We refer the reader to  Appendix A for  the details and provide only the main findings  here.   
%The relation $\rho_r\rho_{\perp} = 1$ no longer holds in 3D, making the three dimensional analysis more complex.

With no loss in generality, see Appendix A,   we  assume   
\beq{53}
\rho_1 > \rho_2 > \rho_3, \text{  with }
\quad \rho_1 > 1,
\quad \rho_3 < 1.
\eeq
The density with the intermediate value, $\rho_2$, may be less than, equal to, or greater than unity.  In order to distinguish these two cases without being specific as to the particular one, we define $\rho_p$ as the density  with value  on the same side of unity as $\rho_2$. 
We assume for the moment  that $\rho_2 \ne 1$; the special case of $\rho_2 = 1$ is discussed separately below. 

The main result is that the physically obtainable material properties can be parameterized in terms of the radial density $\rho_r$, which has a well defined range itself.  Thus, 
${\pmb \phi} \in \Phi_3$ for $ \rho_{rp} \le  \rho_{r}\le  \rho_{r2}$,
where $\rho_{ri}$ are defined in \eqref{-6-6}.  The lower bound is not achieved in practice but is instead set by the value of $ \rho_{r}$ at $r=1$, see   Appendix A. 
The  physically reachable values of the volume fractions, compressibility and density $\rho_\perp$ are therefore  defined through $\rho_r$ as 
 \begin{subequations}\label{62}
\bal{62a}
{\pmb \phi}  &= {\pmb \phi}_{rp}
+ 
\big( 
\frac{\rho_{r}-\rho_{rp} }{\rho_{r2}-\rho_{rp} }
\big) 
( {\pmb \phi}_{r2} - {\pmb \phi}_{rp}), 
\\
C_{*} &= 
C_{*p}
+ 
\big( 
\frac{\rho_{r}-\rho_{rp} }{\rho_{r2}-\rho_{rp} }
\big) 
( C_{*2} - C_{*p}), 
\label{6+}
\\
\rho_\perp &=  \rho_r^{-1}, \quad 
\text{ for  }
%\frac\lambda{1+(\lambda-1)\mu} 
\frac\alpha{1-\beta} 
\le  \rho_{r}\le  \rho_{r2}, 
\label{-77=}
\end{align}
 \end{subequations}
where the critical values $\rho_{ri}$ of the  radial density are defined in \eqref{-6-6},  and the  critical values of the  concentrations ${\pmb \phi}$, and 
compressibilities $C_{*i}$, 
 are 
\beq{51}
{\pmb \phi}_{ri} = 
\left. 
{\pmb \phi}\right|_{\rho_{r}= \rho_{ri}};
\quad
C_{*\, i} = 
\left. 
C_{*}\right|_{\rho_{r}= \rho_{ri}}=  {\pmb C}^T{\pmb \phi}_{ri}.
\eeq
%Note that the definition $\rho_{ri}$ is consistent with that of $\rho_{r3}$ in \eqref{=2}, which in the present context is the special case $\phi_3 = 0$, and hence  $\rho_r =  \rho_{r3}$ as found in \S\ref{sec1}.

\begin{figure}[ht]  %%%%%%%%%%%%%% %%%%%%%%%%%%%% %%%%%%%%%%%%%% %%%%%%%%%%%%%% %%%%%%%%%%%%%%
\begin{center}
 \includegraphics[width=2.5in]{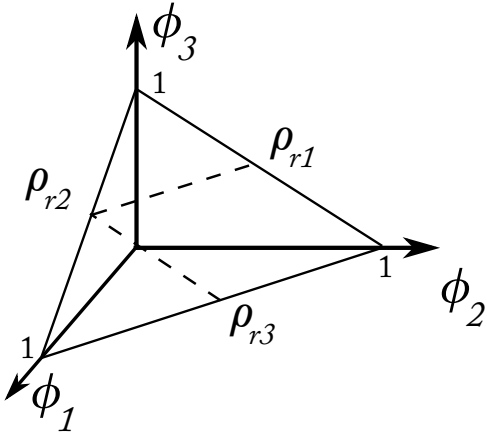} %{phi_map.eps} 
\end{center} 
\caption{The range of ${\pmb \phi}$ for the 3-fluid in the cylindrical configuration.  The dashed lines show the  possible straight line paths as a function of $\rho_{r}$.  In practice, the path begins at some point inside the triangular region $(r=R=1)$ and ends at $\phi_2 = 0$ $(r=r_0, R=R_0)$. }
 \label{fig2}
\end{figure}

Based on the sensitivity analysis in the Appendix, the  radial  density 
 $\rho_r $ has its greatest range, defined by 
 $\Delta \rho_r \equiv \rho_{r2} - \rho_{rp}$, 
  if $\rho_1$ is large, $\rho_2$ is close to unity, and $\rho_3$ is small. 
Thus, 
\beq{-66}
\rho_1 \gg  \rho_2 \approx 1 \gg  \rho_3
\quad
\Rightarrow
\quad
\Delta \rho_r \approx \frac{\rho_1 -1}{1 + \rho_1\rho_3}.
\eeq
The optimal strategy seems to have  three fluids with properties in line with 
\eqref{-66}.  For instance, if $(\rho_1,\rho_2,\rho_3)=(10,1.1,0.01)$, then 
$\Delta \rho_r =8.002$ as compared to   $8.182$ according to \eqref{-66}.
The scalings of \eqref{-66} also imply that the relative magnitudes of the concentrations of the light and heavy fluids are
\beq{=72}
\phi_3 (r) \approx \rho_1\rho_3\, \phi_1  (r).
\eeq

It is also shown in the Appendix that 
\beq{-3}
(C_{*2} - \rho_{r2} )(C_{*p} - \rho_{rp} )\le 0 , 
\eeq
which places another constraint on the choice of the three fluids involving their compressibilities. We next consider these extra degrees of freedom in the context of a special  case of \eqref{-66} which  makes the parameterization simpler.

\subsection{The case of $\rho_2 = 1$ and other limits}

The previous results, in particular the suggested optimal strategy for choosing the densities of the three fluids, suggests that  the results will not depend strongly on $\rho_2 $ if it is close to unity.  It is therefore reasonable to  simply take $\rho_2 =1$, which leads to other simplifications which we now examine. 

The reachable  line in $ \Phi_3$ has one end at the vertex $\phi_2 =1$, and 
\eqref{62} becomes
\begin{subequations}
 \bal{623}
{\pmb \phi}  &= \begin{pmatrix} 0\\ 1 \\0\end{pmatrix} 
+ 
\bigg[ {\pmb \phi}_{r2} - 
\begin{pmatrix} 0\\ 1 \\0\end{pmatrix} 
\bigg]\big( 
\frac{\rho_{r}-1 }{\rho_{r2}-1 }
\big) ,
\\
C_* &= C_2 + 
\big( C_{*2} - C_2\big) 
\big( 
\frac{\rho_{r}-1 }{\rho_{r2}-1 }
\big) ,
\end{align}
\end{subequations}
for the same range of $\rho_{r}$ as in \eqref{62}, and 
with 
\beq{46-}
%\rho_{r2} = \frac{ \rho_1 +\rho_3 } {1+\rho_1 \rho_3 }, \quad 
{\pmb \phi}_{r2} =  \frac{ \rho_{r2} } {\rho_1^2-\rho_3^2 }
\begin{pmatrix}  (1-\rho_3^2) \rho_1 \\0 \\ (\rho_1^2 -1) \rho_3
\end{pmatrix} ,
\quad
C_{*2} =  {\pmb \phi}_{r2}^T 
\begin{pmatrix} C_1 \\0 \\ C_3\end{pmatrix} .
\nonumber
\eeq
The two parameters in the transformation function \eqref{251} simplify, using, 
 \eqref{27}, to  
  \beq{273}
\lambda = S_2, 
 \quad
\mu %= \frac{ C_{*2} - S_2 }{ C_{*2} - \rho_{r2} S_2 }
=1 +  \frac{  	     (\rho_1 -1) (1-\rho_3)(\rho_1-\rho_3) S_2 }  { (1-\rho_3^2) S_1    + (\rho_1^2 -1) S_3 -(\rho_1^2-\rho_3^2)S_2},
\nonumber
 \eeq
 where  $S_i= \rho_i C_i$, $i=1,2,3$. 
  Equations  \eqref{0632}   simplify for
 $\rho_2 = 1$, using  \eqref{-3-}, 
 to give  
 \begin{subequations}\label{036}
 \bal{036s}
R_0  &=    \bigg\{\frac{ S_2 -1}{ s_0^2 S_2 -1} \bigg\}^\frac1{2(1-\mu )}
, 
\quad
r_0 = s_0 R_0, 
 \\
s_0 &=  \bigg\{
 \frac{  (1-\rho_3^2)S_1+(\rho_1^2 -1)S_3} 
 {\rho_1^2 -\rho_3^2 } \bigg\}^{- \frac12 } .
 \end{align}
 \end{subequations} 
We next examine these exact results for some limiting values of the other 3-fluid parameters. 

Both quantities in \eqref{036} should be small.  Based on the assumed density scalings \eqref{-66}, it follows that  $\frac{R_0^2}{r_0^2}$ can be small only if both 
$S_1=$O$(1)$  and $S_3=$o$(1)$.  Under these circumstances,  
%This will be true if $\frac{R_0^2}{r_0^2} \ll 1$. 
eq.  \eqref{-3}, which is now $\big(S_2 -1 \big) 
\big(1- \frac{R_0^2}{r_0^2} \big)
>0$,   requires  $S_2 >1$, and \eqref{273}$_2$ implies in turn that $\mu \approx 0$.  
We therefore have, in addition to 
\eqref{-66} for the densities, that the quantities $S_i$, $i=1,2,3$, should satisfy
$
 S_1=\text{O}(1)$, 
$S_2 > 1$  and $S_3 \ll 1$.
 
\subsubsection{The case $\rho_2=1$, $\rho_1 = \rho_3^{-1}$}
Further simplification results from setting $\rho_3= \rho_1^{-1}$, still with $\rho_1 \gg 1$.  For instance, the volume fractions of phases $1$ and $3$ are equal, 
 \beq{947}
\phi_1 =\phi_3 = \frac 12 (1-\phi_2)=
\frac12 \big(\frac{\rho_{r}-1 }{\rho_{r2}-1 }\big) ,
\eeq
and  $\rho_{r2}$ reduces to $\rho_{r2} = \frac12( \rho_1 + \rho_1^{-1})$.
%Also, \eqref{273} implies that  \beq{3-0-} \frac1{1-\mu} = \big( 1 - \frac{R_0^2}{r_0^2 S_2}\big) \big( 1 + \frac{2\rho_1}{ (\rho_1-1)^2} \big) \approx 1,  \eeq

\subsubsection{Summary}  
 
Based on the analysis above it appears that optimal choices for the properties of the three fluids are 
 \beq{566}
 \rho_1 \gg  \rho_2 = 1 \gg  \rho_3,
\quad  S_1=\text{O}(1), 
\quad S_2 > 1, \quad
S_3 \ll 1,  
 \eeq
 implying $  \lambda = S_2$,  $\mu \approx   0 $. 
Under these circumstances,  \eqref{036} provides the relatively simple approximations
for the values of the inner radius $r_0$, and its  pre-transformed value, $R_0$, 
\begin{subequations}\label{0=-}
\bal{0=-a}
r_0 &\approx \big(1-\frac{ 1}{S_2}\big)^{1/2},
\\
R_0 &\approx \big\{ \big(1-\frac{1}{S_2}\big)(S_3 + S_1 \rho_1^{-2} )
\big\}^{1/2} \ll 1 .
\end{align} 
\end{subequations}
The value of $r_0$ can be made to be close to unity by further requiring 
\beq{4=3}
S_2 \gg 1
\,  \Rightarrow \, 
r_0 \approx 1-\frac{1}{2S_2}, 
\quad 
R_0 \approx   \big( S_3 + S_1 \rho_1^{-2}  
\big)^{1/2}  .
\eeq 
For this range of parameters  the thickness of the physical shell, $1-r_0 \approx  \frac{1}{2S_2}$, depends only on the squared slowness $S_2$, while the image of the inner radius, $R_0$, is dependent on the other two slownesses, and the density $\rho_1$.  While the parameters $r_0$ and $R_0$ are insensitive to the densities $\rho_2 =$O$(1)$ and $\rho_3 =$o$(1)$, the other quantities, such as $C_*$ and $\pmb{\phi}$ can depend on these.  However, if $\rho_3 =1/\rho_1$ then the concentrations of fluids 1 and 3 are everywhere the same. 

 %%%%%%%%%%%%%%%%%%%%%%%%%%%%%%%%%%%%%%%%%%%%%%%%%%%%%%%%%%%%%%%%%%%%%%%
%\section{The three fluid material in $3D$} 

\section{Numerical results}\label{sec6} 

%%%%%%%%%%%%%%%%%%%%%%%%%%%%%%%%%%%%%%%%%%%%%%%%%%%%%%%%%%%%%%%%   TABLE  1
\begin{table}
\caption{\label{tab1}The four cases of 3-fluid material considered.}
\begin{ruledtabular}
\begin{tabular}{ccccccc}
Case  &$\rho_1$  & $\rho_2$ & $\rho_3$  
      &$S_1$     &$S_2$     &$S_3$
 \\
\hline
1&10 & 1 & 0.2 & 1 & 10 & 0.1 
\\
2&10 & 1 & 0.2 & 1 & 10 & 0.01 
\\
3&100 & 1 & 0.02 & 1 & 10 & 0.01 
\\
4&1000 & 1 & 0.002 & 1 & 10 & 0.01 
\\
\end{tabular}
\end{ruledtabular}
\end{table}%%%%%%%%%%%%%%%%%%%%%%%%%%%%%%%%%%%%%%%%%%%%%%%%%%%%%%%%%%%%%%%% 

\subsection{Example of three-fluid shells} 

\begin{figure}[ht] %%%%%%%%%%%%%% %%%%%%%%%%%%%% %%%%%%%%%%%%%% %%%%%%%%%%%%%% %%%%%%%%%%%%%% FIGURE 3
  \begin{center}
  \includegraphics[width=3.4in, height =4in]{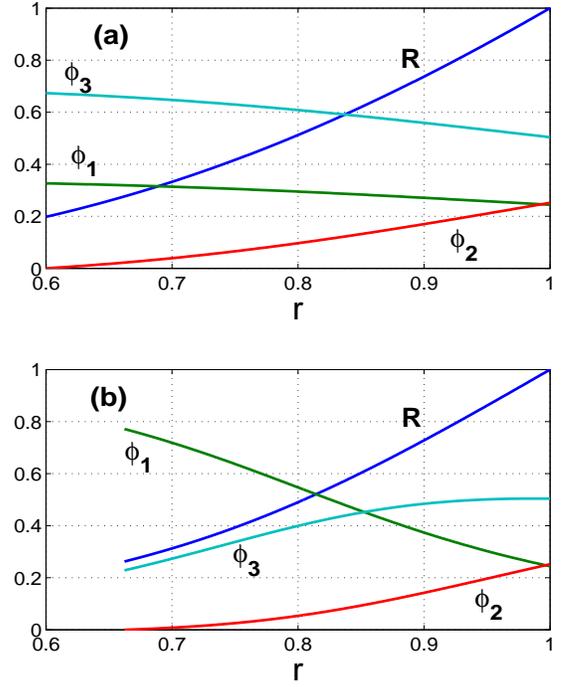}%{f2_2_2D_3D.eps}
  \end{center}
    \caption{(color online) The curves show the concentrations of the three fluids and the radius $R$ as functions of the physical radial coordinate $r$ for the fluid parameters of Case 1 (see Table 1).     (a) the 2D cylindrical configuration;   (b) the 3D spherical shell.}
  \label{fig3}
\end{figure}

\begin{figure}[ht] %%%%%%%%%%%%%% %%%%%%%%%%%%%% %%%%%%%%%%%%%% %%%%%%%%%%%%%% %%%%%%%%%%%%%% FIGURE 4
  \begin{center}
  \includegraphics[width=3.4in, height=4in]{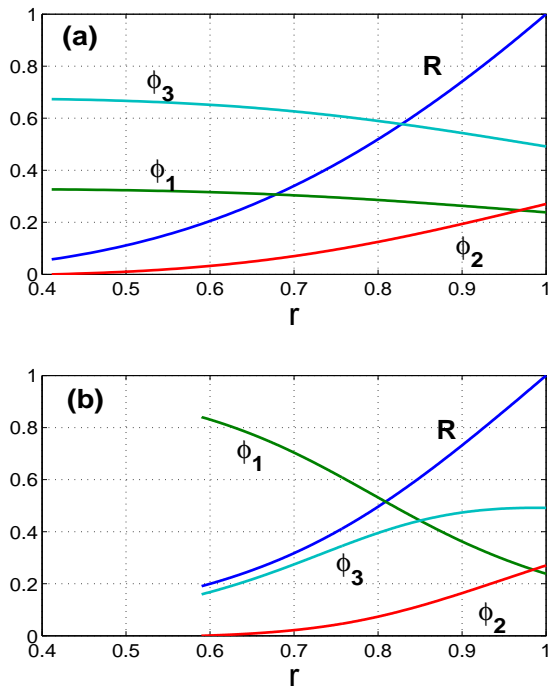}%{f3_2_2D_3D.eps}
  \end{center}
    \caption{(color online) Case 2.  The parameters  are the same as in Fig. \ref{fig3} with the exception that now
$ S_3=0.01 $.  }
  \label{fig4}
\end{figure}

\begin{figure}[ht] %%%%%%%%%%%%%% %%%%%%%%%%%%%% %%%%%%%%%%%%%% %%%%%%%%%%%%%% %%%%%%%%%%%%%% FIGURE 5
  \begin{center}
  \includegraphics[width=3.4in, height =4in]{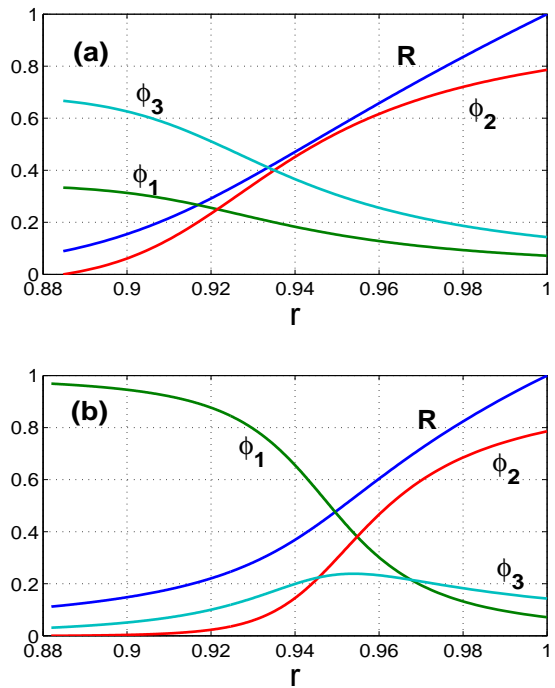}%{f4_2_2D_3D.eps}
  \end{center}  \caption{(color online) Case 3. The parameters  are the same as in Fig. \ref{fig4}   except that 
$ \rho_1= 100 $, $ \rho_3= 0.02$.  }
  \label{fig5}
\end{figure}

\begin{figure}[ht] %%%%%%%%%%%%%% %%%%%%%%%%%%%% %%%%%%%%%%%%%% %%%%%%%%%%%%%% %%%%%%%%%%%%%% FIGURE 6
  \begin{center}
  \includegraphics[width=3.4in, height =4in]{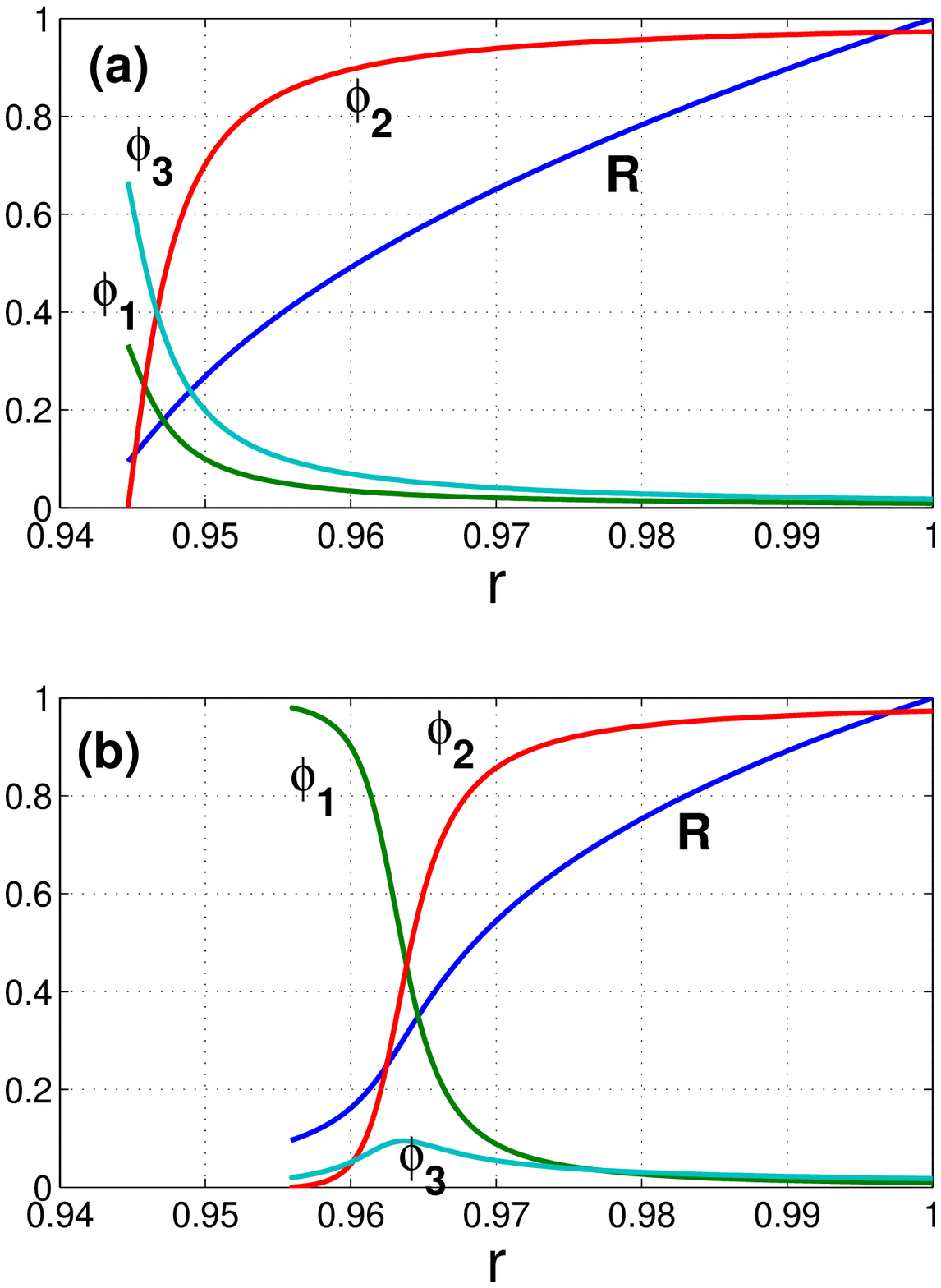}%{f5_2_2D_3D.eps}
  \end{center}
    \caption{(color online) Case 4.  As in Fig. \ref{fig5}   except that now
$ \rho_1= 1000 $, $ \rho_3= 0.002$. }
  \label{fig6}
\end{figure}

The range of possibilities for the 3-fluid metamaterials is extensive  given that there are $3\times 2=6$ independent variables at our disposal.  However, based on the estimates in Section \ref{sec5}, particularly \eqref{566},  it seems reasonable to take $\rho_2=S_1=1$. We further take $\rho_3=2/\rho_1$, in keeping with \eqref{566}. Also, considering  \eqref{4=3} we choose  $S_2=10$, which  leaves two parameters: $\rho_1$ and $S_3$.   Four distinct  3-fluids  are considered according to the four  sets of parameters in Table I %\ref{tab1} 
with different combinations of $\rho_1$ and $S_3$. 
The transformation functions and the concentrations of the three fluid constituents are illustrated in Figs. \ref{fig3}- \ref{fig6}.   The curves $R= R(r)$ illustrate the transformation, 
which maps the original region $R_0 \le R\le 1$ to the physical domain $r_0 \le r\le 1$, 
and the values of the inner radii, $r_0$ and $R_0$, are given in Table II. %\ref{tab2}.  
Note that  $R  \le r$, as expected.    Also, the concentrations for the 2D shells, in  Figs. \ref{fig3}a, \ref{fig4}a, \ref{fig5}a and \ref{fig6}a,  satisfy $\phi_3 \approx 2 \phi_1$, in accordance with \eqref{=72} since $\rho_1\rho_3=2$. 
The most important aspect is the relative values of  $r_0$ and $R_0$, in that it is desirable to have $r_0$  close to unity while $R_0$ should be close to zero.  The value of $r_0$ is smallest in Fig. \ref{fig3} and largest in  Fig. \ref{fig6}, and it appears to increase with $\rho_1$.  In order to obtain a value of $r_0$ close to unity, and in good approximation with  the 
estimate \eqref{4=3}$_1$, it is necessary to have a large value of $\rho_1$, see Figs. \ref{fig5} and \ref{fig6}.   Although only two values of $S_3$ are considered here, numerical experiments indicate that  the value of $R_0$  is more sensitive to this parameter, with   $R_0$ decreasing as $S_3$ is increased.  It is also found that better results, i.e. smaller $R_0$, larger $r_0$, are obtained when $S_2$ becomes very large. For instance, 
$r_0=0.989$, $R_0=0.031$ is obtained in 2D with $\rho_1=S_2=10^3$, $S_3=10^{-3}$. 
 %%%%%%%%%%%%%% %%%%%%%%%%%%%% %%%%%%%%%%%%%% %%%%%%%%%%%%%% %%%%%%%%%%%%%% 
 %%%%%%%%%%%%%% %%%%%%%%%%%%%% %%%%%%%%%%%%%% %%%%%%%%%%%%%% %%%%%%%%%%%%%% 
 %%%%%%%%%%%%%% %%%%%%%%%%%%%% %%%%%%%%%%%%%% %%%%%%%%%%%%%% %%%%%%%%%%%%%% 
 %%%%%%%%%%%%%% %%%%%%%%%%%%%% %%%%%%%%%%%%%% %%%%%%%%%%%%%% %%%%%%%%%%%%%%

\subsection{Discrete layering algorithm}

\begin{figure}[ht]  %%%%%%%%%%%%%% %%%%%%%%%%%%%% %%%%%%%%%%%%%% %%%%%%%%%%%%%% %%%%%%%%%%%%%%
\begin{center}
 \includegraphics[width=2.5in]{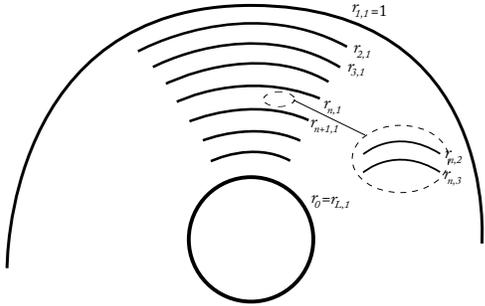}%{circle_layers.eps} 
\end{center} 
\caption{The discrete layering algorithm to reproduce the local homogenization properties of the 3-fluid shell.}
 \label{fig7}
\end{figure}

The inhomogeneous nature of the homogenized material is captured by layering the shell on two scales.  The first scale is a fine layering of $L$ distinct bands defined by the regions between $r_0 < r_1 < r_2 < 
\ldots < r_L = r_{out}=1$.    
The second scale of layering defines three sub-regions between neighboring radii.  Let $r_{n,1} \equiv r_n$, and define
\begin{subequations}\label{04=}
\bal{04=a}
r_{n,m}^d &=r^d_{n,m-1} -   \phi_{m-1}(r_n)  \Delta_n, \quad m=2,3,
\\
\Delta_n &= r_{n}^d - r_{n-1}^d,\quad 
n=1, 2, \ldots ,L, 
\label{04=b}
\end{align}
\end{subequations}
where $\frac\pi 3 (d+1)\Delta_n$ is the area or volume between the inner and outer  radii of the band $[r_{n-1},r_{n}]$. The three regions
$(r_{n,2},r_{n,1}]$ , $(r_{n,3},r_{n,2}]$ and $(r_{n-1,1},r_{n,3}]$  have   fractional volumes $\phi_1(r_n)$, $\phi_2(r_n)$ and $\phi_3(r_n)$ of the band, respectively, and are therefore occupied by the respective fluids, see Fig. \ref{fig7}.   The choice of the ordered set $\{r_n, n=1, 2, \ldots, L-1 \}$ is relatively arbitrary as long as it is finely spaced for large values of $L$. For simplicity we take $\Delta_n $ constant, independent of $n$, in which case $\Delta_n =(1-r_0^d)/L \equiv  \Delta$
and the radii become
\begin{subequations}\label{07=}
\bal{07=a}
r_{n,1}^d &= r_0^d + n\Delta, \quad n=1,2,\ldots ,L, 
\\
r_{n,m}^d &=r^d_{n,m-1} -   \phi_{m-1}(r_{n,1}) \Delta , \quad m=2,3. 
%\\ \Delta &  =(1-r_0^d)/L .  
\label{07=b}
\end{align}
\end{subequations}
%%%%%%%%%%%%%%%%%%%%%%%%%%%%%%%%%%%%%%%%%%%%%%%%%%%%%%%%%%%%%%%%   TABLE  2
\begin{table}
\caption{\label{tab2}Results for the four cases of Table I. %\ref{tab1}. 
$\bar\rho$ is the average density in the shell $r_0\le r \le 1$. 
$\sigma_0$ is the relative value of the total scattering cross section at $kr_0=3$ of a rigid cylinder/sphere surrounded by the 3-fluid shell with 500 layers.  A value of $100\%$ corresponds to the bare rigid target.  }
\begin{ruledtabular}
\begin{tabular}{c|cccc|cccc}
 & & 2D & & &  & 3D & & 
 \\
   &$r_0$ &$R_0$ &$\bar\rho$ &$\sigma_0 (\%)$ 
      &$r_0$ &$R_0$ &$\bar\rho$ &$\sigma_0 (\%)$ 
 \\
\hline
1&0.60 & 0.20 &  3.12  & 25.8  & 0.66 & 0.26 & 5.41& 4.55
\\
2&0.41 & 0.06 &  3.13  & 2.37  & 0.59 & 0.19 & 5.69 &2.20
\\
3&0.88 & 0.09 &  19.17 &0.69  &  0.88 & 0.11 & 57.7 &.033
\\
4& 0.94 & 0.09 & 40.22 & 0.69  & 0.96 & 0.096 & 192 &.012
\\
\end{tabular}
\end{ruledtabular}
\end{table}%%%%%%%%%%%%%%%%%%%%%%%%%%%%%%%%%%%%%%%%%%%%%%%%%%%%%%%%%%%%%%%%   

\subsection{Scattering from a three-fluid shell}

We consider plane wave incidence in the uniform exterior fluid $r>1$, with  time harmonic dependence $e^{-i\omega t}$ (henceforth omitted).   The 3-fluid shell in  $r_0 < r<1$  is defined by the discrete layering algorithm, and is assumed to surround a rigid object of radius $r_0$. 
The scattered pressure is expressed 
\beq{-456}
 p(r,\theta) =  \sum_{n=0}^{\infty} A_n  \psi_n(r ) \Lambda_n(\theta ) , 
\eeq
where $\theta $ is the polar angle with respect to the incident direction,  
 $\{\psi_n(r ), \Lambda_n(\theta )\} =\{  H_n^{(1)}(k r), \cos n \theta\} $ in 2D and $\{  h_n^{(1)}(k r), P_n (\cos \theta) \}  $ in 3D, and 
$k = \omega$ is the nondimensional wavenumber.  
In the shell region the  pressure $p$ and radial velocity $v$ are expressed in modal form
\beq{0220}
\big( p(r,\theta),\, v(r,\theta)\big) =  \sum_{n=0}^{\infty} 
\big( p_n(r),\, v_n(r)\big) \Lambda_n(\theta ), \quad r_0 < r < 1 .
\eeq
  The 
2-vector ${\pmb U}(r) = \big(p_n(r),\, r^{d-1} v_n(r)\big)^T$ satisfies the ordinary differential equation (ODE)
\beq{-=11}
\frac{\dd {\pmb U}}{\dd r} = {\pmb Q}(r) {\pmb U}(r), \quad r_0 < r < 1,
\eeq
where 
\beq{-=12}
 {\pmb Q}(r)  =  \frac{i \omega}{r^{d-1}}
  \begin{pmatrix} 0 & \rho \\ r^{2d-4}\big( \frac{r^2}{K} - \frac{\chi^2}{\omega^2\rho}\big) & 0 \end{pmatrix},
\eeq
and $\chi^2 = n^2, \, n(n+1)$ in 2D and 3D, respectively.   The density $\rho(r)$ and bulk modulus $K(r)$ are piecewise constant,  defined by the 3-fluid material properties at each value of $r$ according to the discrete layering algorithm.  

\subsubsection{Computational scheme}

Three different numerical methods are employed to find the scattered pressure \eqref{-456}: (i) by solving for the matricant; (ii) using a global matrix; and (iii) by solving  the matricant of the homogenized radially dependent anisotropic fluid.  In the first method the matricant \cite{Pease}, or propagator matrix, is found by numerical integration of the matrix equation  $\dd {\pmb M} / \dd r = {\pmb Q}{\pmb M}$ subject to the initial condition ${\pmb M}(r_0) = {\pmb I}$, the 2$\times$2 identity matrix.  Then using the continuity conditions at $r=1$, and the rigid boundary conditions at $r=r_0$, it is possible to express the scattering coefficient $A_n$ in terms of ${\pmb M}(1)$.    Solution (ii) using the global matrix method, e.g.  \cite{Ricks94}, is obtained by creating a large system of simultaneous equations which can be cast as a matrix equation of size $6L$. The third method (ii) is based on the equations of motion of an anisotropic acoustic 
fluid, e.g. \cite{Norris08b},   with radially varying parameters 
$\rho_r$, $ \rho_{\perp}$ and $C_{*}$ given by the exact transformation formulas \eqref{4-4}.   The equations of motion can be transformed into the form \eqref{-=11} with ${\pmb Q} \rightarrow {\pmb Q}_*$ where 
 \beq{-=14}
 {\pmb Q}_*(r)  =  \frac{i \omega}{r^{d-1}}
  \begin{pmatrix} 0 & \rho_r \\ r^{2d-4}\big( \frac{r^2}{K_*} - \frac{\chi^2}{\omega^2\rho_\perp}\big) & 0 \end{pmatrix},
\eeq
and $K_{*}=C_{*}^{-1}$. Using an ODE solver  it is again possible to find the scattering coefficient $A_n$. 
Details of numerical schemes (i) and (iii) will be provided in a forthcoming paper. 

\begin{figure}[ht] %%%%%%%%%%%%%% %%%%%%%%%%%%%% %%%%%%%%%%%%%% %%%%%%%%%%%%%% %%%%%%%%%%%%%% FIGURE 1
  \begin{center}
  \includegraphics[width=3.4in]{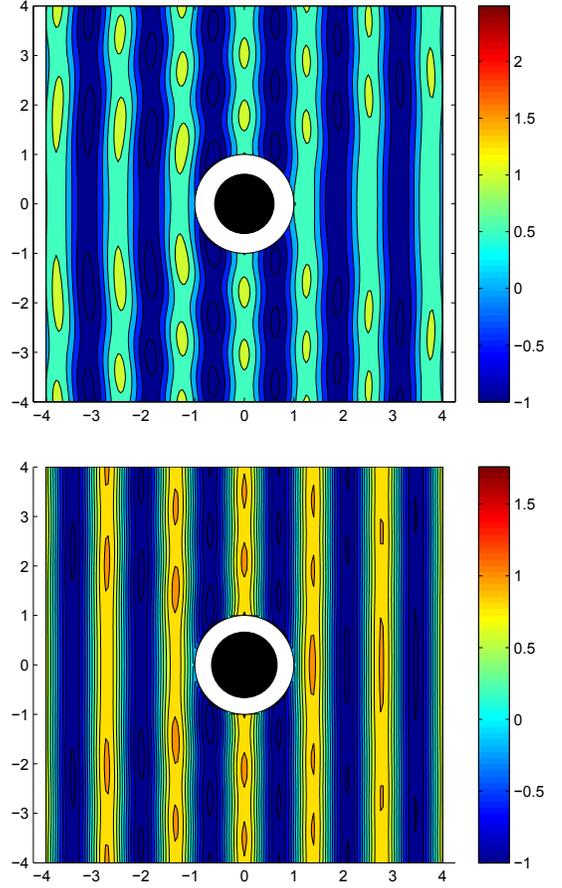}%{fig_1_pressuremap_2D_3D.eps} 
  \end{center}
  \caption{(color online) Case 1.  The magnitude of the  scattered pressure for an incident wave of unit amplitude for the 2D (top) and 3D (bottom) 3-fluid shells.  In each case $kr_0=3$ and $L=500$. The inner dark circular region depicts the rigid target of radius $r_0$, surrounded by the shell of unit outer radius. }
  \label{fig8}
\end{figure} %%%%%%%%%%%%%% %%%%%%%%%%%%%% %%%%%%%%%%%%%% %%%%%%%%%%%%%% %%%%%%%%%%%%%% 
\begin{figure}[ht] %%%%%%%%%%%%%% %%%%%%%%%%%%%% %%%%%%%%%%%%%% %%%%%%%%%%%%%% %%%%%%%%%%%%%% FIGURE 1
  \begin{center}
  \includegraphics[width=3.4in]{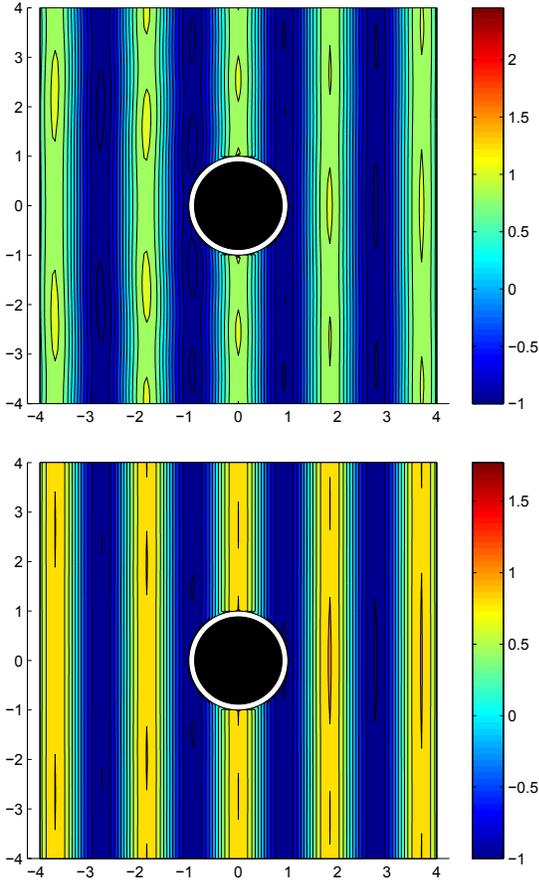}%{fig_2_pressuremap_2D_3D.eps}
  \end{center}
  \caption{(color online) Case 3.  The same as for Figure \ref{fig8}: 2D and 3D simulations are in the upper and lower plots, respectively.   }
  \label{fig9}
\end{figure} %%%%%%%%%%%%%% %%%%%%%%%%%%%% %%%%%%%%%%%%%% %%%%%%%%%%%%%% %%%%%%%%%%%%%% 

\subsubsection{Numerical results}

\begin{figure}[ht] %%%%%%%%%%%%%% %%%%%%%%%%%%%% %%%%%%%%%%%%%% %%%%%%%%%%%%%% %%%%%%%%%%%%%% FIGURE 5
  \begin{center}   % 1.3984 6.8694
  \includegraphics[width=3.4in]{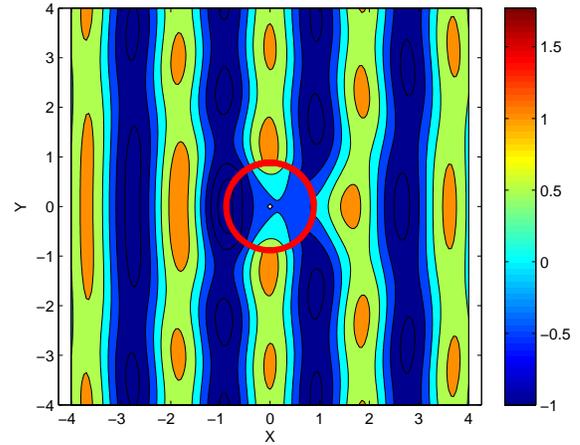}%{fig_5_rigid.eps}
  \end{center}
  \caption{(color online) 3D pressure map solution for a rigid cylinder; $kr_0=3$, $r_0 =.88$.}
  \label{fig10}
\end{figure}

\begin{figure}[ht] %%%%%%%%%%%%%% %%%%%%%%%%%%%% %%%%%%%%%%%%%% %%%%%%%%%%%%%% %%%%%%%%%%%%%% FIGURE 5
  \begin{center}   % 1.3984 6.8694
 \includegraphics[width=3.4in]{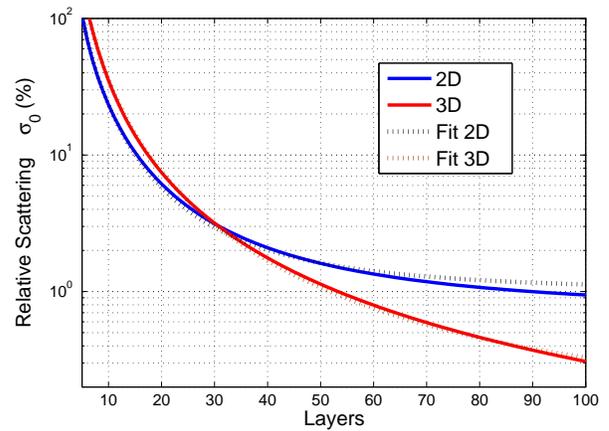}%{fig_5_2D_3D_layers.eps} 
  \end{center}
  \caption{(color online) The relative value of the total scattering cross section for case 3 vs. the number of three-fluid layers in $r_0<r\leq1$.    The dashed lines are curve fits  of the form $aL^b +c$ for $5\leq L \leq 100$.  The parameters $(a,b,c)$ and the  root mean squared error (RMSE) were found to be $ (3716, -2.221,0.9924)$, $0.290$ 
    for 2D, and  $(6435, -2.258,0.1324)$, $0.278$ for 3D.  }
  \label{fig11}
\end{figure}

Figures \ref{fig8} and \ref{fig9} show the magnitude of the  scattered acoustic field for  an incident wave of unit amplitude.  
Since the radius of the object being cloaked changes for each of the four cases of Table I  %\ref{tab1} 
we take the nondimensional characteristic value $kr_0 = 3$ in each scattering simulation.  This allows us to compare the total scattering cross-section between the four cases   even though the values of $r_0$ are different.  Fig. \ref{fig10} shows the response of the bare 3D spherical rigid target based upon case 3 in which $r_0 = .88$.  The total scattering cross-section for the ``cloaked" rigid object  was calculated  using the coefficients $A_n$, and compared with the cross-section for the bare rigid object.  In each case, as Table II %\ref{tab2} 
shows, the relative cross-section is diminished, and for cases 3 and 4 the reduction is significant.  Note that the reduction in target strength is greater in 3D as compared with 2D, in agreement with the general findings of \cite{Norris08b}.  The numerical methods (i) and (ii) were found to be in agreement with one another, and with method (iii) when $L$ is very large.  For instance,  the cross-section found using  method (iii) is $0.3\%$ larger than that of method (i) for the 2D example in Fig. \ref{fig8}.  Finally Fig. \ref{fig11} shows the effect of  the number of layers $L$ on the relative value of the total scattering cross section  for case 3.  A curve fit of the power function $aL^b + c$ shows that for this particular case at this particular frequency the cross-section decreases  as 
$\sigma_0\sim L^{-2.2}$.   More layers provide a better approximation to the homogenized limit, as expected.  For small numbers of layers the layering algorithm used here could be improved using various optimization strategies, but we do not pursue that here. 
%Future improvements in optimization of the layering of the three fluids to result in an effective homogenized medium with fewer layers may significantly improve the results of Fig. \ref{fig11}. 

 %%%%%%%%%%%%%% %%%%%%%%%%%%%% %%%%%%%%%%%%%% %%%%%%%%%%%%%% %%%%%%%%%%%%%% 

 %%%%%%%%%%%%%% %%%%%%%%%%%%%% %%%%%%%%%%%%%% %%%%%%%%%%%%%% %%%%%%%%%%%%%% 

\section{Conclusion}\label{sec7}

The main finding  of this paper is that it is possible to achieve cloaking-like behavior with  as few as three  distinct acoustic fluids.   Using transformation acoustics, we find that for a given set of three fluids the layered shell $r_0<r<1$ is uniquely determined, with the inner radius $r_0$ given by eqs. \eqref{0632} and \eqref{068} for 2D and 3D, respectively.   The shell is made of fine layers of the three fluids with relative concentrations  as a function of $r$ determined  from eqs. \eqref{4-4}, \eqref{40} and \eqref{0345}.  Obviously, 
  the overall effectiveness and the precise form of the layering depends upon the relative densities and compressibilities of the three fluids.  The best results are obtained if one fluid has density equal or close to the background or host fluid density, while the other two densities are much greater and much less than the background value.  Numerical simulations of the scattering from   specific layering realizations confirm the 
   theoretical predictions and show the effect of the finite number of layers. 
Many questions remain as to the  the optimal choice of fluids in general, and what can be achieved using  existing fluids specifically.

\section*{Acknowledgments}
Thanks to an astute referee for suggestions. 
This work was completed with support from the National Science Foundation and from the Office of Naval Research. 

%%%%%%%%%%%%%%%%%%%%%%%%%%%%%%%%%%%%%%%%%%%%%%%%%%%%%%%%%%%%%%%%%%%%%%%%
\appendix
\section{Properties of the 2D 3-fluid material}

\subsection{Density and compressibility}

We begin with the density implications. 
Equation \eqref{40} reduces, using $\rho_r\rho_\perp = 1$,  to give
\beq{43}
\phi_i =  \frac{\rho_i  (\rho_j+\rho_k)  (\rho_r -  \rho_{ri})}
 { \rho_{ri} (\rho_i-\rho_j)(\rho_i-\rho_k)}
,
\quad i\ne j\ne k \ne i,
\eeq
where the critical values of $\rho_r$ are given by \eqref{-6-6}. 
  Based upon the identities \eqref{43}, we note that 
\beq{52}
\left. 
\phi_i\right|_{\rho_{r}= \rho_{ri}}
= 0,
\quad
\left. 
\phi_j\right|_{\rho_{r}= \rho_{ri}}
= \rho_j \rho_{ri} \bigg( \frac{ 1- \rho_k^2 }{\rho_j^2 -\rho_k^2 }\bigg) 
,
\eeq
where $i\ne j\ne k \ne i $. 
The points defined by \eqref{52}  are the intersections of the line  \eqref{43} with the planes 
${\pmb e}_i \cdot {\pmb \phi} = 0$.  
In order to have some ${\pmb \phi} \in \Phi_3$ at least one of the 
intersections must lie on the boundary of $\Phi_3$.  Consider $\rho_{ri}$ of \eqref{-6-6}, then $\phi_j$ and $\phi_k$ must both be positive, which  occurs if and only if  one of $(\rho_j, \rho_k)$ is larger than, and the other is less than, unity.  This gives an important necessary condition: At least one of the three densities is larger than unity, and conversely, at least one must be less than unity.   This condition must hold in addition to the obvious requirement that the three densities are distinct, since otherwise the system \eqref{1} is not solvable. 

Introduce the  density values $\rho_p$, $\rho_m$, $\{p \ne m\} \in \{ 1,3\}$,  such that 
\beq{54}
(\rho_2 -1) (\rho_p -1) > 0, 
\quad 
(\rho_2 -1) (\rho_m -1) < 0,
\eeq
with $2\ne p\ne m \ne 2$. 

We note some other properties of the critical values of the densities:  
\begin{subequations}
\bal{03454}
\rho_{ri} - \rho_{rj} & = \rho_{ri}  \rho_{rj}
\frac{ (\rho_{j} - \rho_{i}) ( 1 - \rho_{k}^2) }
{ (\rho_{i} + \rho_{k})(\rho_{j} + \rho_{k})  }, 
\\
\rho_{ri} - 1 &=  -\rho_{ri}   
\frac{ (\rho_{rj} - 1) ( 1 - \rho_{rk}) }
{ \rho_{rj} + \rho_{rk}  }, 
\end{align}
\end{subequations}
where $i\ne j\ne k \ne i$.  These imply, respectively, that 
$\rho_{r2} > \rho_{rp} > \rho_{rm}$,  and  $\rho_{r2} >1$, $\rho_{rp} >1$ and $\rho_{rm} <1$.  Combining these with the previous inequalities, we surmise  the ordering 
$\rho_{r2} > \rho_{rp} > 1 > \rho_{rm}$.  
Thus, for instance, if $\rho_2 >1$, then the  possible range of 
$\rho_r$ is $\rho_{r1} \le  \rho_{r}\le  \rho_{r2}$.  If 
$\rho_2 <1$ then it is $\rho_{r3} \le  \rho_{r}\le  \rho_{r2}$. 

Any value of $\rho_r$ in the range $\rho_{rp} \le  \rho_{r}\le  \rho_{r2}$ therefore yields a triple of concentration values satisfying ${\pmb \phi} \in \Phi_3$.  
At the upper (lower) value, 
$\rho_{r} = \rho_{r2}$ $( = \rho_{rp})$, the concentration ${\pmb \phi}$ lies on the boundary of the triangle with $\phi_2 = 0$ $(\phi_p = 0)$.  But these limiting values are not necessarily achieved.  Thus, at $r=R=1$ the differential equality \eqref{0345} implies that $\rho_r = \alpha/(1-\beta)$, see eq. \eqref{12}.  This is the practical lower bound on the range of $\rho_r$.  
Equations \eqref{62} and \eqref{51}$_1$ then follow. 

By analogy with equation \eqref{62} for the volume fractions, the effective compressibility of \eqref{69} can be expressed in the form \eqref{6+}.
Alternatively,  eq. \eqref{69} implies $C_{*} = \alpha + \beta  \rho_r $, 
and therefore   we deduce that $\alpha $ and $ \beta$ may be expressed 
\beq{641}
\alpha  = \frac{   \rho_{r2} C_{*\, p}- \rho_{rp} C_{*\, 2}}
{  \rho_{r2}   - \rho_{rp} } ,
\quad
  \beta   = \frac{   C_{*\, 2} - C_{*\, p} }
{  \rho_{r2}   - \rho_{rp} } .
 \eeq
 These lead in turn to explicit expressions for the two parameters that define the transformation function, \eqref{251}, 
\bal{27}
\lambda &= S_1 
\frac{ ( 1-\rho_2)( 1-\rho_3) }{  (\rho_1-\rho_2)(\rho_1-\rho_3) }
+S_2
\frac{  ( 1-\rho_3)( 1-\rho_1) }{  (\rho_2-\rho_3)(\rho_2-\rho_1) }
 \nonumber 
\\ & \quad 
+S_3
\frac{  ( 1-\rho_1)( 1-\rho_2) }{  (\rho_3-\rho_1)(\rho_3-\rho_2) },
 \\
\mu &= \frac{ 
	  \rho_{r1}^{-1} S_1(\rho_2^2-\rho_3^2) 
 +  \rho_{r2}^{-1} S_2(\rho_3^2-\rho_1^2) 
 +  \rho_{r3}^{-1} S_3(\rho_1^2-\rho_2^2) }
 { S_1 (\rho_2^2-\rho_3^2)
 +S_2 (\rho_3^2-\rho_1^2)
 +S_3 (\rho_1^2-\rho_2^2)},
 \nonumber 
\end{align}
where  $S_i =\rho_i C_i$, $i=1,2,3$.
 
 \subsubsection{Sensitivity}

The reachable range of $\rho_r$ is, 
from \eqref{0345}, $\rho_{rp} < \rho_{r} < \rho_{rp}+\Delta \rho_r$ where
\beq{03433}
\Delta \rho_r \equiv
\rho_{r2} - \rho_{rp} =
\frac{ (\rho_{p} - \rho_{2}) ( 1 - \rho_{m}^2) }
{ (1+\rho_{2}  \rho_{m})(1+\rho_{p}  \rho_{m})  }.
\eeq
Hence, 
%\begin{subequations}
\bal{34-}
\frac{\partial \Delta \rho_r}{\partial \rho_2}
 &= -\frac{(1 - \rho_{m}^2) }{ (1+\rho_{2}  \rho_{m})^2},
  \nonumber 
\\
\frac{\partial \Delta \rho_r}{\partial \rho_p}
 &= \frac{(1 - \rho_{m}^2) }{ (1+\rho_{p}  \rho_{m})^2},
 \\
\frac{\partial \Delta \rho_r}{\partial \rho_m}
 &= - \frac{ (\rho_{p} - \rho_{2}) ( 1 + \rho_{m}^2) 
 (\rho_{2}+\rho_{p}+2\rho_{2}\rho_{p}\rho_{m})}
{ (1+\rho_{2}  \rho_{m})^2(1+\rho_{p}  \rho_{m})^2  }
. \nonumber 
\end{align}
%\end{subequations}
If $p=1$ these are, respectively, $<0$, $>0$, $<0$.  Conversely, 
if $p=1$ they are $>0$, $<0$, $>0$.   Hence, whether $p=1$ or $p=3$ it is clear that $\Delta \rho_r $ is greatest if 
$\rho_1$ is large, $\rho_2$ is close to unity, and $\rho_3$ is small.

\subsection{Transformation function}

\subsubsection{Necessary conditions for the three-fluid parameters}

Since $R(1)=1$ at the outer radius  $r=1$, we have 
\beq{12}
R'(1) = \rho_r(1) = C_*(1) = \frac{\alpha }{1-\beta },  
\eeq
that is, 
\beq{121}
 \rho_r(1) = C_*(1) = 
 \frac{\rho_{r2}C_{*p} - \rho_{rp}C_{*2} }
 {\rho_{r2}- \rho_{rp} +C_{*p} -C_{*2} }.
\eeq
But we require that
$
\rho_{rp} \le \rho_r(1) \le \rho_{r2}$,  
or, since $\rho_{r2}- \rho_{rp} >0$, 
\beq{-2}
\begin{split}
 \frac{ \rho_r(1) - \rho_{rp} }{\rho_{r2}- \rho_{rp}}&= \frac{ C_{*p}-  \rho_{rp} } {\rho_{r2}- \rho_{rp} +C_{*p} -C_{*2} } >0 
  ,  
  \\
  \frac{\rho_{r2}-\rho_r(1) }{\rho_{r2}- \rho_{rp}}&= 
  \frac{ \rho_{r2} -C_{*2} } {\rho_{r2}- \rho_{rp} +C_{*p} -C_{*2} } >0 .
  \end{split}
\eeq
Hence, \eqref{-3} must hold,  
%Therefore, we have two cases: 
%\beq{-48} \begin{split} (i) & \quad C_{*p} > \rho_{rp} \text{  and  } C_{*2} < \rho_{r2}, \text{ or} \\  (ii) & \quad C_{*p} < \rho_{rp} \text{  and  } C_{*2} > \rho_{r2}.   \end{split} \eeq
or, explicitly
\bal{-3-}
%\big(\frac{C_{*2}}{\rho_{r2} } -1 \big) \big(\frac{C_{*p}}{\rho_{rp} } -1 \big) =
&\big(S_1 \frac{1-\rho_3^2}{\rho_1^2 -\rho_3^2} + S_3 \frac{\rho_1^2 -1}{\rho_1^2 -\rho_3^2}-1\big)\times 
\nonumber \\
& 
\big(
S_2 \frac{1-\rho_m^2}{ \rho_2^2 -\rho_m^2 } + S_m \frac{\rho_2^2 -1}{ \rho_2^2 -\rho_m^2 }-1 \big) <0. 
\end{align}

\subsubsection{The case $\rho_2 = 1$}
In this case the  $p/m$ distinction is unnecessary since 
\beq{554}
\rho_{r1}=\rho_{r3}=1\text{ for }\rho_2 = 1, 
\, 
\Rightarrow \, 
{\pmb \phi} \in \Phi_3 \, \text{ for  }
1 \le  \rho_{r}\le  \rho_{r2}.
\nonumber 
\eeq
This implies that the reachable quantities reduce to \eqref{623}. 

\section{The two fluid material}\label{sec3}

\subsection{General  theory}
%%%%%%%%%%%%%%%%%%%%%%%%%%%%%%%%%%%%%%%%%%%%%%%%%%%%%%%%%%%%%%%%%%%%%%%%%%%%%%
The 2-fluid version of eq. \eqref{1} is 
\begin{equation} \label{known}
\begin{pmatrix} 1 & 1   \\ \rho_1 & \rho_2     \\ \frac{1}{\rho_1} & \frac{1}{\rho_2}   \end{pmatrix} \begin{pmatrix} \phi_1 \\ \phi_2   \end{pmatrix} = \begin{pmatrix} 1 \\ \rho_r \\ \rho_{\perp}^{-1} \end{pmatrix}. 
\end{equation}
This implies that the concentrations are 
\beq{011}
\phi_1  =  \frac{\rho_r- \rho_2} {\rho_1 - \rho_2}, 
\qquad
\phi_2  = \frac{\rho_1- \rho_r} {\rho_1 - \rho_2},
\eeq
and the densities $\rho_r$, $\rho_\perp$ are related by the compatibility condition for \eqref{known}, 
\beq{-44}
\rho_r + \rho_1\rho_2 \rho_{\perp}^{-1} =
\rho_1 + \rho_2 .
\eeq
The effective compressibility, which follows from \eqref{011} and the third relation in  \eqref{4-3},  satisfies 
\begin{equation} \label{-33}
(\rho_1 - \rho_2) C_* +(C_2-C_1) \rho_r = \rho_1 C_2- \rho_2 C_1
.  
\end{equation} 
Equation \eqref{011} provides relations for the volume fractions in terms of the radial inertia $\rho_r$. One can also interpret eqs. \eqref{-44}  and \eqref{-33} as defining 
$\rho_{\perp}$ and $C_*$, respectively, in terms of $\rho_r$.  Therefore, all parameters in the two-fluid material can be defined by a single quantity, in this case  $\rho_r$.

However, in order to relate the two-fluid material to a transformation it is necessary that there exists a function $R$ which satisfies the three differential identities \eqref{4-4}.  Substitution of these into eqs. \eqref{-44} and \eqref{-33} gives a pair of equations which  can be considered as algebraic equations in two unknowns:  $R'$ and $R/r$. Solutions for both of these quantities can be found in terms of the two-fluid properties $\rho_1,\rho_2, C_1,C_2$, but the solutions are not of practical interest.  The reason is that the constant values of $R'$ and $R/r$ that are found, say $R'=a$, $R/r = b$, must be equal, leading to trivial cases.  
%We will return to this in Appendix B after we have discussed the 3-fluid material.  
The main conclusion from the study of the $N=2$ case is that the 2-fluid material is overly restrictive.

\subsection{A special case of a uniform 2-fluid material} 

While it is not possible for the 2-fluid material to reproduce a transformation material, it is possible to make some interesting uniform fluids with anisotropic inertia. 
The idea is to seek constant values of $\rho_r$, $\rho_{\perp}$ and $C_*$ which also match to the exterior fluid in $r>1$. This requires that $R=1$ at $r=1$. Enforcement of  \eqref{4-4} then requires the three parameters in the left vector be equal to $R'$. 
Substituting into eqs. \eqref{-44}  yields 
\beq{-6}
\rho_r = \rho_{\perp}^{-1} =  C_* =     %\frac{\rho_1+\rho_2}{1+\rho_1\rho_2} \equiv  
\rho_{r3} .  
\eeq
The volume fractions follow from \eqref{011} as 
\beq{=2}
 \phi_1 =  \rho_1   \rho_{r3} \bigg( \frac{1-\rho_2^2}{\rho_1^2 - \rho_2^2}\bigg),
 \quad
 \phi_2 =  \rho_2   \rho_{r3} \bigg( \frac{1-\rho_1^2}{\rho_2^2 - \rho_1^2}\bigg),
 %\phi_i =  \rho_i   \rho_{r3} \bigg(  \frac{1-\rho_j^2}{\rho_i^2 - \rho_j^2}  \bigg), \, \, i=1,2, \, \, j\ne i, 
 \eeq
 which  are both positive if and only if 
$(1-\rho_1)(1-\rho_2) <0$.   
%Therefore, with no loss in generality, we  specify \beq{=3} \rho_1 > 1 > \rho_2. \eeq
The one remaining condition,  for the compressibility,  implies using \eqref{-33} and \eqref{-6} that  the two compressibilities must be related such that
  \beq{=42}
  C_1  \rho_1 (1- \rho_2^2) +C_2 \rho_2 ( \rho_1^2-1)
 =
 \rho_1^2 - \rho_2^2.
 \eeq

The anisotropic fluid \eqref{-6} is defined by  the parameter $\rho_{r3}=\rho_{r3}(\rho_1, \rho_2)$, and is composed of volume fractions  
$\phi_i=\phi_i(\rho_1, \rho_2)$ of fluid $i=1,2$. 
Denote  any pair satisfying the  relation \eqref{=42} as $C_i = C_i( \rho_1, \rho_2)$, $i=1,2$.  
It is interesting to note  that these functions are invariant under  the interchange  
$\{ \rho_1, \rho_2, \phi_1,\phi_2,C_1,C_2\}\rightarrow 
\{ \rho_2^{-1}, \rho_1^{-1}, \phi_2,\phi_1,C_2,C_1\}$. 
% \beq{-4}  \rho_r(\frac1{\rho_2} , \frac1{\rho_1}) = \rho_r(\rho_1, \rho_2),  \quad  \phi_i(\frac1{\rho_2} , \frac1{\rho_1})  = \phi_j(\rho_1, \rho_2),  \quad  C_i(\frac1{\rho_2} , \frac1{\rho_1})  = C_j(\rho_1, \rho_2), \, i\ne j .  \eeq

\subsubsection{Examples}
If, for instance,  $C_1  \rho_1 = C_2  \rho_2$ then \eqref{=42} implies that 
$C_1  \rho_1 = C_2  \rho_2 =1$.  Both fluids have the same wave speed as the background fluid.  They differ only in their impedances,  
which in this case are $z_i = \rho_i = C_i^{-1}$, $i=1,2$.

Conversely, if  $C_1  /\rho_1 = C_2  /\rho_2$ then \eqref{=42} implies that 
$C_1  /\rho_1 = C_2  /\rho_2 =1$.  The two  fluids have the same acoustic impedance as the background fluid, and  differ only in their wave speeds,  
which   are $c_i = \rho_i^{-1} = C_i^{-1}$, $i=1,2$.

\subsection{A two and a half fluid material }

As a case intermediate between the strictly 2-fluid and 3-fluid cases,  consider the 2D case, for which $(r/R)R' = \rho_r$, see \eqref{4-4}$_1$.    It follows from \eqref{-44}$_2$, i.e.  $\rho_r = \rho_\perp^{-1}$,  that $\rho_r = \rho_{r3}$, a constant.  Taking into account the boundary condition $R(1) = 1$, the unique mapping is 
\beq{0523}
R(r) = r^{ \rho_{r3}}. 
\eeq
Equation \eqref{-33}  combined with \eqref{4-4}$_3$ then implies 
\beq{-23-}
(1-\rho_2^2)S_1 + (\rho_1^2 - 1)S_2 = (\rho_1^2 -\rho_2^2 ) r^{ 2(\rho_{r3}-1)}.
\eeq
This cannot be satisfied if the two fluids have properties independent of $r$.  However, if we still require that the densities are fixed, but the compressibilities could vary with $r$, then 
\eqref{-23-} suggests that a mapping can be realized if one or both  $S_1$, $S_2$ are such that the equality holds for some range of $r$. 
It is well known that adding a small concentration of bubbles to a liquid results in an increase in the compressibility   without significant change in the effective density.  Hence, it might be possible, in principle if not in practice, to add a third fluid whose only role is to enhance compressibility.  In this sense it is half of a fluid, since its inertial properties are not used. 

%Since $(\rho_1 -1)(\rho_3 -1)$ must be negative, we take, with no loss in generality 
%$\rho_1 >1>\rho_3$.  A large value of $\rho_{r3}$ is achieved if 
%$\rho_1 \gg1$, $\rho_3 \ll 1$, in which case \eqref{-23-} becomes 
%\beq{-24-}
%\rho_1^{-2}S_1 +  S_2 \approx  r^{ 2(\rho_{r3}-1)}.
%\eeq 

%%%%%%%%%%%%%%%%%%%%%%%%%%%%%%%%%%%%%%%%%%%%%%%%%%%%%%%%%%%%%%%%%%%%%%%

%\bibliography{../bib/thermoelastic}

\begin{thebibliography}{9}
\providecommand{\natexlab}[1]{#1}
\providecommand{\url}[1]{\texttt{#1}}
\expandafter\ifx\csname urlstyle\endcsname\relax
  \providecommand{\doi}[1]{doi: #1}\else
  \providecommand{\doi}{doi: \begingroup \urlstyle{rm}\Url}\fi

\bibitem[Norris(2009)]{Norris09}
A.~N. Norris.
\newblock Acoustic metafluids.
\newblock \emph{J. Acoust. Soc. Am.}, 125\penalty0 (2):\penalty0 839--849,
  2009.
\newblock \doi{10.1121/1.2817359}.

\bibitem[Pendry et~al.(2006)Pendry, Schurig, and Smith]{Pendry06}
J.~B. Pendry, D.~Schurig, and D.~R. Smith.
\newblock Controlling electromagnetic fields.
\newblock \emph{Science}, 312\penalty0 (5781):\penalty0 1780--1782, June 2006.
\newblock \doi{10.1126/science.1125907}.

\bibitem[Cummer and Schurig(2007)]{Cummer07}
S.~A. Cummer and D.~Schurig.
\newblock One path to acoustic cloaking.
\newblock \emph{New J. Phys.}, 9\penalty0 (3):\penalty0 45+, March 2007.
\newblock \doi{10.1088/1367-2630/9/3/045}.

\bibitem[Chen and Chan(2007)]{Chen07}
H.~Chen and C.~T. Chan.
\newblock Acoustic cloaking in three dimensions using acoustic metamaterials.
\newblock \emph{Appl. Phys. Lett.}, 91\penalty0 (18):\penalty0 183518+, 2007.
\newblock \doi{10.1063/1.2803315}.

\bibitem[Norris(2008)]{Norris08b}
A.~N. Norris.
\newblock Acoustic cloaking theory.
\newblock \emph{Proc. R. Soc. A}, 464:\penalty0 2411--2434, 2008.
\newblock \doi{10.1098/rspa.2008.0076}.

\bibitem[Torrent and S\'{a}nchez-Dehesa(2008)]{Torrent08b}
D.~Torrent and J.~S\'{a}nchez-Dehesa.
\newblock Acoustic cloaking in two dimensions: a feasible approach.
\newblock \emph{New J. Phys.}, 10\penalty0 (6):\penalty0 063015+, June 2008.
\newblock \doi{10.1088/1367-2630/10/6/063015}.

\bibitem[Schoenberg and Sen(1983)]{Schoenberg83}
M.~Schoenberg and P.~N. Sen.
\newblock Properties of a periodically stratified acoustic half-space and its
  relation to a {B}iot fluid.
\newblock \emph{J. Acoust. Soc. Am.}, 73\penalty0 (1):\penalty0 61--67, 1983.
\newblock \doi{10.1121/1.388724}.

\bibitem[Pease(1965)]{Pease}
M.~C. Pease.
\newblock \emph{Methods of Matrix Algebra}, 172--176.
\newblock Academic Press, New York, 1965.

\bibitem[Ricks and Schmidt(1994)]{Ricks94}
D.~C. Ricks and H.~Schmidt.
\newblock A numerically stable global matrix method for cylindrically layered
  shells excited by ring forces.
\newblock \emph{J. Acoust. Soc. Am.}, 95\penalty0 (6):\penalty0 3339--3349,
  1994.

\end{thebibliography}
%\bibliographystyle{unsrtnat}%unsrt}%plain}%abbrvnat}%plain}%doipubmed}%harvard}%natbib}
%\end{document}

\end{document}